\documentclass[acmsmall,screen]{acmart}

\AtBeginDocument{%
  }

\setcopyright{cc}
\setcctype{by}
\acmDOI{10.1145/3808123}
\acmYear{2026}
\acmJournal{PACMSE}
\acmVolume{3}
\acmNumber{FSE}
\acmArticle{FSE116}
\acmMonth{7}
\acmSubmissionID{fse26mainb-p293-p}
\received{2026-02-10}
\received[accepted]{2026-03-24}

\usepackage{multirow}
\usepackage{tabularx}
\usepackage{subcaption}
\usepackage{graphicx}
\usepackage{tcolorbox}
\usepackage{amsmath}
\usepackage{amsfonts}
\usepackage{fancyvrb}
\usepackage{xcolor}
\usepackage{wrapfig}

\newtcolorbox{observation}[1][]{
    colback=yellow!10!white,
    colframe=black,
    rounded corners,
    arc=5pt,
    boxrule=1pt,
    left=5pt, right=5pt, top=1pt, bottom=1pt,
    before skip=7pt, after skip=7pt,
    #1
}

\newtcolorbox{takeaway}[1][]{
    colback=blue!10!white,
    colframe=black,
    rounded corners,
    arc=5pt,
    boxrule=1pt,
    left=5pt, right=5pt, top=1pt, bottom=1pt,
    before skip=7pt, after skip=7pt,
    #1
}

\usepackage{listings}
\usepackage{caption}
\tcbuselibrary{listings}

\newcounter{codecounter}[subsection]

\definecolor{terminalBg}{RGB}{240,235,255}
\definecolor{terminalGreen}{RGB}{34,139,34}
\definecolor{terminalRed}{RGB}{205,92,92}
\definecolor{terminalGray}{RGB}{128,128,128}

\newtcblisting{pythoncode}{
    listing only,
    listing options={
        language=Python,
        basicstyle=\ttfamily\footnotesize,
        commentstyle=\color{terminalGreen},
        keywordstyle=\color{terminalGreen}\bfseries,
        stringstyle=\color{terminalRed},
        numberstyle=\tiny\color{terminalGray},
        breaklines=true,
        showstringspaces=false,
        morekeywords={def, return, if, else, pass, raise}
    },
    colback=terminalBg,
    colframe=black!20,
    coltitle=black,
    arc=3mm,
    boxrule=0.5pt,
    left=1pt, right=1pt, top=1pt, bottom=1pt,
    before skip=1\baselineskip,
    after skip=1\baselineskip
}

\newcommand{\revised}[1]{\textcolor{black}{#1}}
\newenvironment{revisions}{\color{black}}{}

\usepackage{enumerate}
\usepackage{enumitem}
\usepackage{needspace}

\begin{document}

\title[Reward-Free Code Alignment from Pretrained or Fine-Tuned LLM: Unpacking the Trade-offs for \dots]{Reward-Free Code Alignment from Pretrained or Fine-Tuned LLM: Unpacking the Trade-offs for Code Generation}

\author{Sanjeepan Sivapiran}
\affiliation{%
  \institution{York University}
  \country{Canada}
}

\author{Gias Uddin}
\affiliation{%
  \institution{York University}
  \country{Canada}
}

\newcommand{\subsubsubsection}[1]{\paragraph{#1}}

\renewcommand{\shortauthors}{Sanjeepan Sivapiran and Gias Uddin}

\begin{abstract}
Large Language Model (LLM) alignment trains an LLM using preference data to produce outputs that better \revised{meet established quality standards (e.g., to mitigate toxic or biased responses in text generation, or to enforce coding best practices in code generation)}.
While LLM alignment techniques are studied for non-coding tasks, we know little about their usefulness for coding tasks. Intuitively, for code generation tasks using LLMs, alignment techniques could help with coding solutions that better \revised{adhere to established software engineering standards}, such as more secure code, supporting better coding practices, etc. However, it is unclear whether LLM code alignment could support both functional requirements (producing executable, correct code) and non-functional requirements (code readability, style, maintainability). It is also unknown whether alignment for a code LLM should begin with base pretrained version or the finetuned (i.e., instruction-tuned) version of the LLM. In this paper, we offer insights on the above two research questions by conducting an empirical study. We studied five state-of-the-art (SOTA) LLMs using two widely used LLM alignment techniques: Direct Preference Optimization (DPO) and BoNBoN. For each training record, we created a preference pair as accepted and rejected instances by using the SelfCodeAlign pipeline. DPO and BoNBoN are reward-free models, i.e., they eliminate the need for multiple reward scores for output preferences. We tuned each LLM using the two alignment techniques in two settings: pretrained and finetuned versions of an LLM. We evaluated functional requirements using four SOTA benchmarks (HumanEval+, MBPP+, EvalPerf, EvoEval) and non-functional requirements using the \revised{CODAL benchmark, which evaluates code quality across five dimensions derived from software engineering practices}. We find that pretrained-to-aligned pathways achieve larger improvements in the aligned variant over its pretrained variant (CodeLlama-7b: +75\% non-functional, Llama3-8b: +42\% functional). But the pretrained variant is generally less accurate than its finetuned variant. However, finetuned-to-aligned offers smaller performance improvements or, in some cases, degradation in the aligned variant than its finetuned variant. This means that while the base pretrained version is less accurate than its base finetuned variant, alignment reduces the performance gap between pretrained and finetuned variants. Non-functional requirements improve more consistently than functional requirements via alignment. Based on these findings, we provide nine recommendations to guide alignment for code LLMs.
\end{abstract}



\maketitle

\section{Introduction}

Large Language Models (LLMs) have demonstrated remarkable capabilities in code generation, with systems like ChatGPT \cite{openai2024gpt4technicalreport}, Claude \cite{anthropic2024claude}, and Gemini\cite{comanici2025gemini} showing the potential to assist developers in programming tasks \cite{jiang2024surveylargelanguagemodels, guo2024deepseek, wang2023codet5+, luo2023wizardcoder, muennighoff2023octopack, wei2023magicoder}. However, their generated code could contain logical flaws, performance inefficiencies, security vulnerabilities, and may not properly meet user requirements \cite{chen2021evaluating, yang2024robustnesssecurityprivacyexplainability}. Traditional approaches to improve code generation have relied primarily on fine-tuning techniques that train pretrained models to follow specific coding instructions \cite{wei2024selfcodealign, ma2024llamoco}. However, LLMs still struggle to consistently meet the strict requirements demanded by software engineering practices, particularly regarding both functional correctness and non-functional code quality \cite{almonte2025automated, ullrich2025requirements}.

LLM alignment has emerged as a methodology for modifying output behavior using paired examples of preferred and non-preferred responses \cite{rafailov2023direct, christiano2017deep, wang2024comprehensive, ouyang2022training, stiennon2020learning, deng2025less}. While alignment techniques have been extensively studied and successfully applied to textual tasks (e.g., to mitigate toxic or biased responses) \cite{wang2024comprehensive, zhang-etal-2024-knowledgeable}, their application to code generation remains relatively unexplored.


\definecolor{lightgreen}{RGB}{220,252,231}
\definecolor{darkgreen}{RGB}{22,101,52}
\definecolor{lightred}{RGB}{254,226,226}
\definecolor{darkred}{RGB}{127,29,29}
\definecolor{lightblue}{RGB}{240,244,248}
\definecolor{darkblue}{RGB}{37,99,235}

\begin{figure}[h]
\centering
\begin{subfigure}[t]{0.48\textwidth}
\centering
\textbf{Non-Coding Task Alignment}
\vspace{0.5em}
\fcolorbox{gray!30}{white}{\begin{minipage}{0.95\textwidth}
\fcolorbox{darkblue}{lightblue}{\begin{minipage}{0.95\textwidth}
\small\textbf{User:} Write a product review for a smartphone.
\end{minipage}}
\vspace{0.3em}
\small\textbf{ ALIGNED RESPONSE:}\\
\fcolorbox{darkgreen}{lightgreen}{\begin{minipage}{0.95\textwidth}
\small\textit{This smartphone offers solid performance with a 6.1-inch display and 12MP camera. Battery life lasts a full day with moderate use. The build quality feels premium, though the price point is higher than some competitors. Consider your budget and specific needs before purchasing.}
\end{minipage}}
\vspace{0.3em}
\small\textbf{ MISALIGNED RESPONSE:}\\
\fcolorbox{darkred}{lightred}{\begin{minipage}{0.95\textwidth}
\small\textit{This is the BEST phone ever made! You MUST buy it immediately! It's perfect in every way and has zero flaws. Everyone who doesn't own this phone is making a terrible mistake!}
\end{minipage}}
\end{minipage}}
\vspace{0.5em}
\small\textit{Focus: Helpfulness, honesty, avoiding manipulation}
\subcaption{Non-coding task alignment}
\label{subfig:noncoding}
\end{subfigure}
\hfill
\begin{subfigure}[t]{0.48\textwidth}
\centering
\textbf{Coding Task Alignment}
\vspace{0.5em}
\fcolorbox{gray!30}{white}{\begin{minipage}{0.95\textwidth}
\fcolorbox{darkblue}{lightblue}{\begin{minipage}{0.95\textwidth}
\small\textbf{User:} Write a function to handle user passwords.
\end{minipage}}
\vspace{0.3em}
\small\textbf{ ALIGNED RESPONSE:}\\
\fcolorbox{darkgreen}{lightgreen}{\begin{minipage}{0.95\textwidth}
\footnotesize\texttt{%
import hashlib, secrets\\[0.1em]
\\[0.1em]
def hash\_password(password):\\[0.1em]
\hspace*{4ex}salt = secrets.token\_hex(16)\\[0.1em]
\hspace*{4ex}hashed = hashlib.pbkdf2\_hmac(\\[0.1em]
\hspace*{8ex}'sha256', password.encode(),\\[0.1em]
\hspace*{8ex}salt.encode(), 100000)\\[0.1em]
\hspace*{4ex}return salt + hashed.hex()%
}
\end{minipage}}
\vspace{0.3em}
\small\textbf{ MISALIGNED RESPONSE:}\\
\fcolorbox{darkred}{lightred}{\begin{minipage}{0.95\textwidth}
\footnotesize\texttt{%
def store\_password(password):\\[0.1em]
\hspace*{4ex}with open('passwords.txt', 'a') as f:\\[0.1em]
\hspace*{8ex}f.write(f"\{password\}{\textbackslash}n")\\[0.1em]
\hspace*{4ex}return password%
}
\end{minipage}}
\end{minipage}}
\vspace{0.5em}
\small\textit{Focus: Security, best practices, avoiding vulnerabilities}
\subcaption{Coding task alignment}
\label{subfig:coding}
\end{subfigure}
\caption{Examples of aligned vs. misaligned LLM responses across different task domains}
\label{fig:alignment_examples}
\end{figure}

Code generation presents unique alignment challenges distinct from general text generation, as illustrated in Figure \ref{fig:alignment_examples}. The left panel \ref{subfig:noncoding} shows non-coding alignment, where aligned responses provide balanced, honest smartphone reviews versus misaligned responses using manipulative language to push purchases, emphasizing helpfulness and avoiding manipulation. The right panel \ref{subfig:coding} demonstrates coding alignment, where aligned responses implement secure password hashing with industry-standard algorithms versus misaligned responses dangerously storing passwords in plain text. Here, alignment prioritizes security and coding best practices over simple correct-only functional solutions. This distinction reveals how alignment objectives must be fundamentally adapted to address domain-specific requirements: while general text generation focuses on truthfulness and user welfare, software engineering alignment must additionally ensure code security, maintainability, and adherence to established programming standards, spanning both functional correctness and critical non-functional qualities.

However, practitioners face two critical choices when applying alignment to code generation: 
\begin{enumerate}
    \item Should code alignment be applied for both functional and non-functional requirements?
    \item Should code alignment begin with pretrained or finetuned version of an LLM?
\end{enumerate}

In this paper, we present an empirical study to guide these two choices. We evaluate five state-of-the-art LLMs (Meta-Llama-3-8B, Qwen2.5-Coder-7B, CodeLlama-7b, deepseek-coder-1.3b, and deepseek-coder-7b) along with their instruction-tuned variants using two alignment techniques: DPO and BoNBoN. We examine both pretrained-to-aligned and finetuned-to-aligned pathways, generating separate preference datasets for functional and non-functional requirements. Models are evaluated on established functional benchmarks (HumanEval+, MBPP+, EvalPerf, EvoEval) and the CODAL benchmark for non-functional requirements.

Our findings reveal distinct patterns from textual alignment tasks \cite{jindal2024balancing}. Unlike textual alignment where technique effectiveness follows general patterns \cite{kotha2023understanding}, code alignment demonstrates task-type dependency and model family-specific preferences. Pretrained-to-aligned models achieve substantial relative improvements (CodeLlama-7b: +75\% non-functional, Llama3-8b: +42\% functional) but from weaker baselines, while finetuned-to-aligned models provide higher absolute performance with minimal improvement. Additionally, we observe model family-specific technique preferences (Meta-Llama performs better with DPO, deepseek models with BoNBoN) and early failure indicators (SFT-stage degradation predicts final alignment failure) that contrast with the generalizable patterns in textual alignment.

Further, our findings show that code alignment should prioritize non-functional over functional requirements due to substantially different success patterns. Non-functional alignment achieved consistent improvements across all models (CodeLlama-7b: +75\%, deepseek-coder-7b: +27.8\%, Qwen2.5-Coder-7B: +15.7\%), while functional alignment ranged from significant success (Llama3-8b: +42\%) to catastrophic failure (CodeLlama-7b: -40\%). Non-functional requirements respond reliably regardless of model architecture, whereas functional requirements exhibit high model dependence with substantial degradation risk. These contrasting success patterns between requirement types are different from textual alignments, where alignment effectiveness remains consistent \cite{jindal2024balancing}.

\begin{takeaway}
\textbf{Summary of Major Findings}: 
The alignment of finetuned LLMs improved the LLM performance in 51\% times for functional coding tasks and 53\% times in non-functional coding tasks. However, the average relative performance improvement was modest (4.9\%) for functional tasks, while it was more pronounced (10.6\%) for non-functional coding tasks. Overall, alignment from finetuned rather than alignment from pretrained offered better performance. The relative performance gain was more significant in pretrained models after alignment, but those aligned pretrained models still were inferior to their finetuned or the aligned finetuned variants. 
\end{takeaway}
\section{Background}
\subsection{LLM Alignment Pathways}



The stability-plasticity dilemma, originally introduced in neuroscience and continual learning research, describes a trade-off phenomenon between a model's ability to retain previously learned knowledge (stability) and its capacity to acquire new information (plasticity) \cite{kirkpatrick2017overcoming}. In neural networks, this manifests as catastrophic forgetting, where models lose performance on previously learned tasks when trained on new data. This phenomenon is particularly relevant to LLM fine-tuning scenarios, where continued training can lead to performance degradation on previously tuned capabilities \cite{luo2025empiricalstudycatastrophicforgetting}. The challenge becomes more complex when considering alignment processes, as models must balance preserving existing functional capabilities while adapting to new preference data.

Recent studies have demonstrated the complex relationship between model starting points and alignment effectiveness. Kotha et al. \cite{kotha2023understanding} show that instruction-tuning fundamentally alters model behavior by suppressing the ability to learn from examples provided in prompts, while Jindal et al. \cite{jindal2024balancing} find that continuous pre-training of instruction-tuned models results in catastrophic forgetting of instruction-following abilities. However, systematic comparisons between pretrained-to-aligned versus instruction-tuned-to-aligned pathways remain limited, particularly for domain-specific applications like code generation.

These findings highlight how LLM alignment effectiveness fundamentally depends on the starting model state. Pretrained base models possess broad language understanding but lack task-specific instruction-following capabilities, demonstrating high plasticity for adaptation but low stability in preserving behaviors \cite{kirkpatrick2017overcoming}. Conversely, instruction-tuned models have undergone supervised fine-tuning on instruction-response pairs, developing structured response patterns and task-following abilities at the cost of reduced plasticity for further adaptation.

This difference becomes particularly important when considering domain-specific alignment. For textual tasks, alignment typically focuses on subjective qualities like helpfulness and harmlessness, where pretrained models require extensive alignment to develop appropriate response patterns while instruction-tuned models need refinement of existing capabilities. Coding tasks present unique challenges, involving both objective functional requirements (correctness, compilation) and subjective non-functional requirements (readability, style). Code-specialized models possess domain-specific knowledge that affects alignment responsiveness differently than general language models.

To address these challenges and systematically compare the trade-offs between stability and plasticity, we consider two distinct alignment pathways.

\begin{itemize}
    \item \textbf{The pretrained-to-aligned pathway} starts from pretrained model. This approach leverages high plasticity but begins from lower task-specific performance. 
    \item \textbf{The finetuned-to-aligned pathway} begins with instruction-tuned models and applies the same training sequence, starting from higher absolute performance.
\end{itemize}

\subsection{Studied LLM Alignment Techniques}
\begin{figure}[htbp]
    \centering
    \begin{subfigure}{0.38\textwidth}
        \centering
        \includegraphics[width=\textwidth]{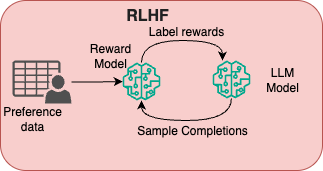}
        \caption{Reinforcement learning from human feedback}
        \label{fig:image1}
    \end{subfigure}
    \hspace{0.1\textwidth}
    \begin{subfigure}{0.25\textwidth}
        \centering
        \includegraphics[width=\textwidth]{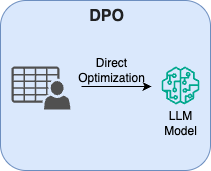}
        \caption{Direct Preference Optimization}
        \label{fig:image2}
    \end{subfigure}
    \caption{RLHF vs DPO}
    \label{fig:rlhf_vs_dpo}
    
\end{figure}

LLM alignment employs several techniques to train models using preference data to produce outputs that better reflect human values and preferences. These approaches can be categorized into two main types: reward-based (RLHF \cite{ouyang2022training}, RLAIF \cite{lee2023rlaif}) and reward-free approaches (DPO \cite{rafailov2023direct}, KTO \cite{ethayarajh2024kto}, ORPO \cite{hong2024orpo}, RPO\cite{yin2024relative}) \cite{xu2024dpo}.

Reward-based techniques include Reinforcement Learning from Human Feedback (RLHF), which represents the foundational approach utilizing a two-stage process, as shown in Figure \ref{fig:rlhf_vs_dpo}(a), where a reward model is first trained on human preference data, followed by policy optimization using Proximal Policy Optimization (PPO) to guide model outputs toward preferred responses \cite{christiano2017deep, ouyang2022training}.

Reward-free techniques, including Direct Preference Optimization (DPO) \cite{rafailov2023direct} and Best-of-N BoN (BoNBoN) \cite{gui2024bonbon}, eliminate reward model dependency by reformulating alignment as a single-objective optimization problem. As shown in Figure \ref{fig:rlhf_vs_dpo}(b), this approach offers computational efficiency through single-stage training, improved stability by avoiding reinforcement learning complexities, and comparable or better empirical performance while using lower resources as discussed in \cite{rafailov2023direct}.

We selected DPO and BoNBoN for our experiments based on their complementary strengths. DPO is well-suited for resource-constrained scenarios and domain-specific applications like code generation, where preference patterns are objective and learnable through classification. BoNBoN achieves theoretically optimal win-rate maximization while eliminating inference-time sampling costs \cite{gui2024bonbon}, making it ideal for deployment scenarios where efficiency is crucial. Together, DPO's classification approach and BoNBoN's data distribution-matching approach provide comprehensive coverage of alignment techniques while maintaining computational feasibility across our experiments.

    


\subsubsection{Direct Preference Optimization (DPO)}
DPO eliminates the reward model requirement by directly optimizing preferences, as illustrated in Figure \ref{fig:rlhf_vs_dpo}(b) \cite{rafailov2023direct}. Unlike RLHF's two-stage process (preference data → reward model → reinforcement learning), DPO uses a single-stage approach where preference data directly trains the final model through maximum likelihood optimization. This reformulation treats alignment as a classification problem, training models to distinguish between preferred and rejected responses without separate reward estimation, as demonstrated in Figure \ref{fig:alignment_examples}(b). The streamlined approach reduces computational overhead while achieving better performance compared to PPO-based RLHF \cite{DBLP:journals/corr/SchulmanWDRK17, christiano2017deep, stiennon2020learning}.

\subsubsection{BoNBoN}

\begin{figure}[htbp]
    \centering
    \begin{subfigure}{0.35\textwidth}
        \centering
        \includegraphics[width=\textwidth]{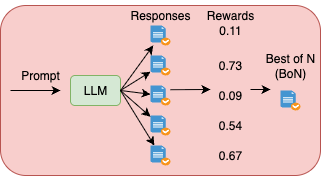}
        \caption{Best-of-N}
        \label{fig:image1}
    \end{subfigure}
    \hspace{0.1\textwidth}
    \begin{subfigure}{0.35\textwidth}
        \centering
        \includegraphics[width=\textwidth]{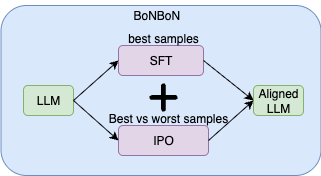}
        \caption{BoNBoN}
        \label{fig:image2}
    \end{subfigure}
    \caption{Best-of N and BoNBoN}
    \label{fig:bon_vs_bonbon}
\end{figure}

Best-of-N sampling addresses alignment by generating multiple responses to each prompt and selecting the highest-scoring option, as shown in Figure \ref{fig:bon_vs_bonbon}(a). While this approach produces high-quality outputs, it requires generating n samples at inference time, creating substantial computational overhead.

BoNBoN (Best-of-N BoN) eliminates this inference cost by training models to directly mimic the best-of-n distribution \cite{gui2024bonbon}. As illustrated in Figure \ref{fig:bon_vs_bonbon}(b), BoNBoN employs a dual training strategy: supervised fine-tuning (SFT) on the best responses from n-sample generations, combined with Iterative Preference Optimization (IPO) \cite{pang2024iterative} training that contrasts the best versus worst responses from the same sample sets. This approach enables the final aligned model to internalize the selection process that best-of-n sampling performs during inference, achieving comparable quality benefits without the computational overhead.

\subsection{Related Work}



While alignment techniques have proven effective for general text generation, their application to code generation remains underexplored \cite{miao2024aligning}. Code generation presents unique challenges requiring both functional correctness and adherence to software engineering standards. Most existing research \cite{zhang2024codedpo, miao2024aligning} focuses on functional requirements using benchmarks like HumanEval \cite{chen2021evaluating}, MBPP \cite{austin2021program}, Classeval \cite{du2023classeval}, CoderEval \cite{yu2024codereval}, and SWE-bench \cite{jimenez2023swe}, while non-functional requirements like readability, style, and maintainability are relatively underexplored \cite{singhal2024nofuneval, almonte2025automated, 10.1145/3736407}. Recent work has begun addressing this gap with benchmarks like CODAL \cite{10.1145/3736407} evaluating non-functional requirements across multiple dimensions.

Instruction tuning \cite{zhang2024gpt4roi} has emerged as standard practice for improving code generation capabilities \cite{ma2024llamoco, li2023instructcoder, tsai2024code, ahmad2025opencodeinstruct}, with instruction-tuned models consistently outperforming base models. However, the impact of different starting points (base vs instruction-tuned) for subsequent alignment processes remains largely unexplored in code generation domains \cite{gondara2025small, springer2025overtrained}.

The stability-plasticity dilemma from continual learning research \cite{kirkpatrick2017overcoming} describes the trade-off between retaining learned knowledge (stability) and acquiring new information (plasticity). This phenomenon, show as catastrophic forgetting in neural networks \cite{ramasesh2021effect}, becomes particularly relevant during LLM alignment, where models must balance preserving existing capabilities while adapting to preference data \cite{luo2025empiricalstudycatastrophicforgetting}. No prior work has examined these trade-offs specifically for code generation alignment or provided guidance for selecting optimal alignment pathways based on functional versus non-functional requirements.

\revised{While foundational work on RLHF for general text generation \cite{ouyang2022training} and reinforcement learning for reasoning \cite{guo2025deepseek} has demonstrated alignment feasibility, neither compared the effects of different starting points (pretrained vs. finetuned) on code generation, nor addressed the distinct behavior of functional versus non-functional requirements under alignment.}

Furthermore, to the best of our knowledge, no prior work has examined the stability-plasticity trade-offs specific to code generation alignment, nor provided guidance for practitioners on selecting optimal alignment pathways based on their specific requirements. This work addresses these gaps by providing the first comprehensive comparison of pretrained-to-aligned versus fine-tuned-to-aligned pathways, with explicit consideration for both functional and non-functional code requirements.

\section{Methodology}
\label{section:Methodology}

We answer three research questions (RQ):
 
\begin{enumerate}[label=\textbf{RQ\arabic{*}.}]
    \item To what extent does preference-based alignment enhance the functional correctness and non-functional properties of base pre-trained code generation models?
    \item How does an initial instruction-tuning phase alter the effectiveness of subsequent preference-based alignment on both the functional correctness and non-functional requirements of code generation models?
    \item What are the fundamental trade-offs in functional and non-functional outcomes when choosing between a pretrained-to-aligned and a finetuned-to-aligned pathway, and how does the starting point affect the risks and efficacy of different alignment techniques?
\end{enumerate}


Our study consists of two steps. In the first step, preference datasets for each model are generated using a modified Selfcodealign pipeline \cite{wei2024selfcodealign} (See Section \ref{sec:creation-of-pref-datasets}).

Since our research questions involve both functional and non-functional requirements, we generate the dataset twice for each model, modifying the prompt each time to reflect the different requirements. This dual dataset generation approach ensures that we capture preferences for both types of requirements separately. The modification of prompts allows us to target specific aspects of code quality that align with our research objectives.


For the alignment process in both pathways, we train the model with supervised fine-tuning, where only the chosen response is used as the target response. We train all models in this step for approximately 120k steps using 100\% of the available data. \revised{During pilot experiments, we observed that beyond 120k iterations, evaluation loss plateaued while training loss continued decreasing. This led to the decision to select 120k steps (2 full epochs).} Additionally, 120k steps provides sufficient training iterations to ensure convergence while avoiding overfitting. This supervised fine-tuning step establishes a solid foundation for subsequent alignment training.

\begin{wraptable}{l}{0.5\textwidth}
\centering
\caption{Coding Task Categories}
\label{tab:codingtasks} 
\small
\setlength{\tabcolsep}{3pt}
\begin{tabular}{|l|p{2.5cm}|c|c|}
\hline
\textbf{ID} & \textbf{Description} & \textbf{Type} & \textbf{Benchmark} \\
\hline
F1 & Code generation & Func & HumanEval+ \\
\hline
F2 & Problem solving & Func & MBPP+ \\
\hline
F3 & Performance & Func & EvalPerf \\
\hline
F4 & Complex reasoning & Func & EvoEval \\
\hline
NF1 & Instruction following & Non-Func & CODAL \\
\hline
NF2 & Readability & Non-Func & CODAL \\
\hline
NF3 & Complexity & Non-Func & CODAL \\
\hline
NF4 & Style adherence & Non-Func & CODAL \\
\hline
NF5 & Code explanation & Non-Func & CODAL \\
\hline
\end{tabular}
\end{wraptable}
\revised{Regarding the risk of catastrophic forgetting when re-applying SFT to instruction-tuned models in the FTA pathway, following Luo et al. \cite{luo2025empiricalstudycatastrophicforgetting}, we note that SFT on preference data alleviates rather than induces forgetting. Our results confirm this: SFT-stage degradation occurred in both PTA and FTA pathways (O8), indicating alignment incompatibility rather than pathway-specific forgetting, and models stable through SFT generally succeeded in alignment (O10).}

After supervised fine-tuning is complete, we proceed to the alignment step. For direct preference optimization (DPO) \cite{rafailov2023direct}, we use 100\% of the available data with preferred and dispreferred columns for training. Similar to the SFT, we train for 120k iterations. We use the SFT-trained model as the base in this step to carry over the SFT learnings to the final model. We follow the same approach for BoNBoN \cite{gui2024bonbon} alignment as well.

\subsection{Studied Coding Tasks}

\setlength{\intextsep}{0pt}
\begin{wraptable}{r}{0.4\textwidth}
\centering
\caption{Dataset Statistics}
\label{tab:dataset_stats}
\small
\begin{tabular}{|l|c|c|}
\hline
\textbf{Model} & \textbf{Type} & \textbf{Samples} \\
\hline
\multirow{2}{*}{Llama-3-8B} & Func & 64,804 \\
\cline{2-3}
 & Non-Func & 68,249 \\
\hline
\multirow{2}{*}{Qwen2.5-7B} & Func & 61,823 \\
\cline{2-3}
 & Non-Func & 72,128 \\
\hline
\multirow{2}{*}{CodeLlama-7B} & Func & 70,435 \\
\cline{2-3}
 & Non-Func & 72,732 \\
\hline
\multirow{2}{*}{DeepSeek-1.3B} & Func & 32,768 \\
\cline{2-3}
 & Non-Func & 31,330 \\
\hline
\multirow{2}{*}{DeepSeek-7B} & Func & 58,345 \\
\cline{2-3}
 & Non-Func & 57,538 \\
\hline
\end{tabular}
\end{wraptable}

While LLMs can generate syntactically correct code, meeting comprehensive software requirements presents a challenge. Production environments demand code that not only executes correctly but also adheres to industry best practices, security protocols, and maintainability. Generated code often lacks proper error handling, coding conventions, or produces functional but inefficient, unreadable solutions \cite{yang2024robustnesssecurityprivacyexplainability}.

This creates a significant gap between code that is correct and code that is production ready. To address this challange, we need alignment tuning designed specifically for coding tasks. Code alignment is different from regular text generation as it must handle both technical correctness and quality standards that change depending on the project and organization.

We organize these requirements into two main types: functional and non-functional. We test both types using standard benchmarks, as shown in Table~\ref{tab:codingtasks}. This approach allows us to measure how well alignment methods work across all aspects of software engineering standards.

Table ~\ref{tab:codingtasks} organizes our evaluation framework into nine distinct tasks spanning both requirement types. Functional tasks (F1-F4) assess objective code correctness across four established benchmarks, progressing from basic generation to complex reasoning scenarios. Non-functional tasks (NF1-NF5) evaluate qualitative aspects through the comprehensive CODAL benchmark, covering the standard software engineering best practices.

\subsubsection{Functional Requirements in Coding Tasks}
\begin{wrapfigure}{l}{0.6\textwidth}
    \includegraphics[width=0.6\textwidth]{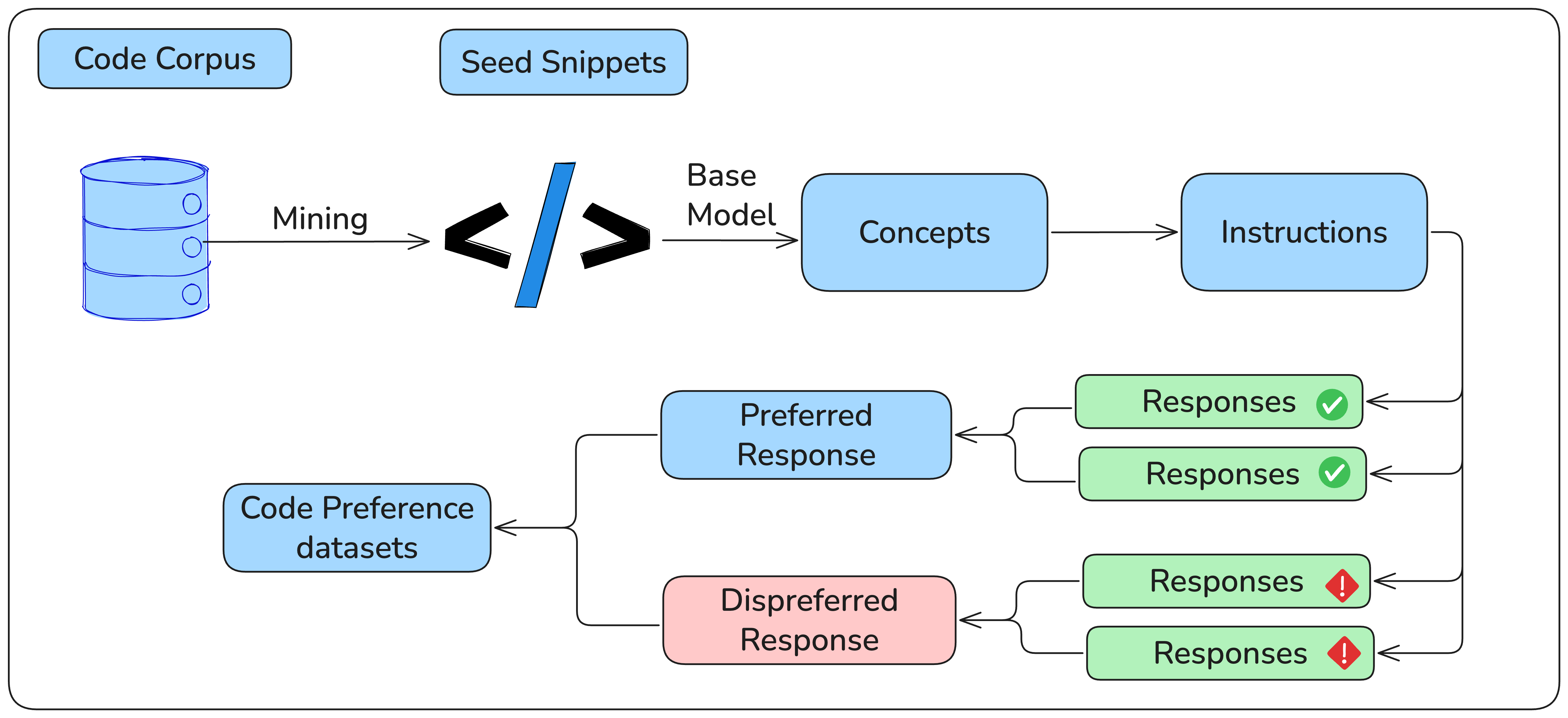}
    \caption{Data Generation Workflow}
    \label{fig:end_to_end_workflow}
\end{wrapfigure}
Functional requirements specify what the code must accomplish to be executable and produce correct outputs. This includes generating syntactically correct, compilable code in target programming languages that follows established standards, incorporates proper error handling (Figure \ref{code:error}), implements security practices, and includes performance optimization.

\subsubsection{Non-functional Requirements}
Non-functional Requirements primarily include instruction following, code explanation, code complexity and efficiency, code readability (Figure \ref{code:readablity})
, and coding style as defined in \cite{10.1145/3736407}. These requirements ensure the generated code meets broader software engineering standards and practices.  \revised{These five dimensions are derived from established software engineering practices rather than direct developer elicitation, representing benchmark-defined proxies for code quality.}

This categorization allows for a more comprehensive evaluation of code LLM alignment across both correctness aspects (functional) and quality aspects (non-functional).

\subsection{Creation of Preference Datasets}\label{sec:creation-of-pref-datasets}
\begin{wrapfigure}{r}{0.4\textwidth}
\vspace{-1\baselineskip}
\begin{pythoncode}
# With proper error handling
def remove_duplicates(lst):
    if not isinstance(lst, list):
        raise TypeError("Input must be a list")
    if not lst:
        return []
    return list(dict.fromkeys(lst))
# Without error handling
def remove_duplicates(lst):
    return list(dict.fromkeys(lst))  # Crashes on None input
\end{pythoncode}
\caption{Error handling}
\label{code:error}
\end{wrapfigure}
Our methodology integrates prior work on LLM alignment and code generation, notably SelfCodeAlign \cite{wei2024selfcodealign} and BoNBoN \cite{gui2024bonbon}. We utilize SelfCodeAlign to generate fully transparent training preference datasets for Direct Preference Optimization (DPO) and BoNBoN LLM alignment techniques, as shown in Figure \ref{fig:end_to_end_workflow}.

\textbf{SelfCodeAlign} proposes a pipeline to generate fully transparent instruction tuning datasets for supervised fine-tuning. In this work, we modify the pipeline to generate preference datasets for LLM alignment. The process begins with seed functions generated from the Stack V1 source code dataset \cite{kocetkov2022stack3tbpermissively}. Next, using a selected base model, coding concepts are extracted. Using these coding concepts, instructions are generated to reconstruct the original code snippet. The same base model then generates multiple responses for each instruction.

These responses are executed and filtered to include only passing responses in the instruction tuning dataset. We modify this pipeline by randomly selecting one passing response as the chosen response and one failing non-empty response as the rejected response. This approach enables us to build preference datasets for each model. Note that we generate preference datasets separately for each model because LLMs learn more efficiently when training data follows the same distribution as the base model's data distribution \cite{wei2024selfcodealign}.

\revised{For non-functional requirements, we generate separate preference datasets for each dimension using dimension-specific prompts. We applied similar modifications across all five evaluation dimensions, ensuring balanced representation (e.g., a 60k non-functional dataset contains 12k samples per dimension). Our results validate this approach: code-specialized models achieved consistent improvements across all five dimensions (O3), and different dimensions showed distinct response patterns (O6), suggesting the preference pairs capture meaningful quality differences.}

Using the above approach, we generated datasets for each model. Table \ref{tab:dataset_stats} shows the models and the number of preference instances for each model. 

\subsection{Model Selection}
Our study evaluates five base open source models along with their corresponding instruction-tuned versions, for a total of 10 models. The selected models include Meta-Llama-3-8B \cite{grattafiori2024llama}, Qwen2.5-Coder-7B\cite{hui2024qwen2}, CodeLlama-7b \cite{roziere2023code}, deepseek-coder-1.3b \cite{guo2024deepseek}, and deepseek-coder-7b \cite{guo2024deepseek}. These are paired with their respective instruction-tuned \cite{zhang2024gpt4roi} versions: Meta-Llama-3-8B-Instruct, Qwen2.5-Coder-7B-Instruct, CodeLlama-7b-Instruct-hf, deepseek-coder-1.3b-instruct, and deepseek-coder-7b-instruct-v1.5.

\begin{wrapfigure}{r}{0.55\textwidth}
\vspace{-1.5\baselineskip}
\begin{pythoncode}
# Readable implementation
def longest_common_prefix(strings):
    if not strings:
        return ""
    
    shortest_string = min(strings, key=len)
    
    for index, char in enumerate(shortest_string):
        for string in strings:
            if string[index] != char:
                return shortest_string[:index]
    
    return shortest_string
    
# Unreadable implementation  
def lcp(s):
    if not s:return ""
    m=min(s,key=len)
    for i,c in enumerate(m):
        for st in s:
            if st[i]!=c:return m[:i]
    return m
\end{pythoncode}
\vspace{-10pt}
\caption{Code Readability}
\label{code:readablity}
\vspace{-1.5\baselineskip}
\end{wrapfigure}
Our model selection prioritizes diversity among selected models while maintaining a small resource footprint. The selection consists of different sizes ranging from 1.3B to 8B parameters, providing insight into how model scale affects alignment effectiveness. We include both general-purpose models (Meta-Llama-3-8B) and code-specialized models (Qwen2.5-Coder, CodeLlama, deepseek-coder), enabling us to investigate whether domain specialization influences alignment outcomes. While other open source variants exist, our selection focuses on models that are: (1) actively maintained with stable releases, (2) widely adopted as benchmarks, and (3) representative of distinct architectures. This diversity allows us to investigate how different architectures affect the effectiveness of alignment pathways while ensuring computational feasibility through our focus on smaller models.

\subsection{Evaluation Benchmarks}

\subsubsection{Evaluation on Functional Benchmarks}

\textbf{HumanEval+ / MBPP+} are functi-onal-level coding benchmarks. HumanEval+ extends the original HumanEval \cite{chen2021evaluating} with 164 instances and 80x more test cases, while MBPP+ enhances MBPP \cite{austin2021program} with 378 instances and 35x more testing coverage. We use evalplus \cite{NEURIPS2023_43e9d647} for standardized evaluation with Pass@1 score.

\textbf{EvalPerf} assesses LLM-generated code efficiency alongside HumanEval+ and MBPP+. This 118-instance benchmark \cite{NEURIPS2023_43e9d647} measures the percentage of solutions meeting reference efficiency.

\textbf{Evoeval} \label{sec:evoeval} provides contamination-free assessment through systematically modified problems. The benchmark contains 828 problems across seven categories, evaluated using Pass@1 score \cite{xia2024top}.

\subsubsection{Evaluation on Non-Functional Benchmarks}

\textbf{CODAL} \cite{10.1145/3736407}  is a non-functional requirements benchmark with 500 problems across five dimensions: instruction-following, code explanation, complexity/efficiency, readability, and coding style. Using LLM-as-judge methodology, it provides 0-10 ratings per dimension and aggregate scores.

\subsection{Evaluation Metrics}
The results in the section \ref{sec:results} are reported as relative improvement, which is defined as \\ $\Delta_{rel,i} = \frac{P_{final,i} - P_{baseline,i}}{P_{baseline,i}} \times 100\%$ \\ where $P_{final,i}, P_{baseline,i}$ are aligned/baseline Pass@1 (functional) or CODAL scores (non-functional). \\
\textbf{Pass@1:}
Percentage of problems solved correctly on first attempt (HumanEval/+, MBPP+, EvoEval). \\
\textbf{CODAL Score:}
LLM judge rates responses 1-10 across 5 \revised{code quality dimensions} (instruction-following, explanation, efficiency, readability, style). Individual \revised{dimension} score $S_p$ averages ratings across 100 problems; overall score averages across \revised{dimensions}. $r_{p,i} \in \{1, \ldots, 10\}$ = LLM judge rating for problem $i$ in \revised{dimension} $p$. $p \in \{1, \ldots, 5\}$ = \revised{code quality dimensions}. We have $S_p = \frac{1}{100} \sum_{i=1}^{100} r_{p,i}$ and $\text{CODAL Score} = \frac{1}{5} \sum_{p=1}^{5} S_p$.

\subsection{Threats to Validity}

Limited model coverage (five base models, 1.3B-8B parameters) may not represent larger models exceeding 70B parameters, which could exhibit emergent properties \cite{wei2022emergent} and could alter alignment effectiveness \cite{muckatira2024emergent}. Utilzing those models, however, will require significant GPU resources. Nevertheless, our findings and study setups could guide companies with bigger models working on code LLMs.
Our evaluation emphasizes general-purpose code generation tasks, potentially missing specialized programming categories like class-level generation or bug fixes that may exhibit different quality requirements. However, alignment objectives for such tasks are also not standerdized, i.e., we may not need alignment for such tasks anyway.
Construct validity concerns arise from our operational definition of alignment through functional and non-functional requirements, which may not encompass all dimensions practitioners consider essential, such as robustness to adversarial conditions or ethical compliance. 

\revised{Additionally, the CODAL benchmark dimensions represent software engineering practices rather than directly validated developer priorities; while these dimensions align with recognized code quality standards, they serve as proxies for code quality rather than developer preferences. All preference pairs in our study are model-specific and self-generated, following SelfCodeAlign's finding that models learn more efficiently from matching distributions \cite{wei2024selfcodealign}. While this maximizes within-model alignment effectiveness, it may encode model-specific stylistic biases rather than broadly generalizable quality patterns. Cross-model preference transfer remains an important direction for future work.}
\section{ Results}
\label{sec:results}

\subsection{Performance Comparison between Pretrained-to-Aligned (RQ1)} 

We analyze how alignment techniques (DPO, BoNBoN) affect pretrained code LLMs across functional (RQ 1.1) and non-functional (RQ 1.2) coding tasks as detailed in Table  \ref{tab:codingtasks}. 

\subsubsection{Functional Coding Tasks in Pretrained Models (RQ 1.1)}

Preference-based alignment produces highly variable functional improvements across coding tasks F1-F4, with some models achieving higher relative improvement despite starting from significantly lower absolute performance baseline scores compared to instruction-tuned variants. Results are shown in Table \ref{tab:evalplus_combined} for F1-F3 tasks and Table \ref{tab:evoeval_combined} for F4 comprehensive evaluation, and all results are reported in Pass@1 scores below.

\subsubsubsection{\textbf{Success Cases}}

Llama3-8b demonstrates the strongest improvements across all functional coding tasks, with basic code generation (F1) improving 42\% (0.372→0.530) with DPO and 47\% on HumanEval+ with DPO/BoNBoN, algorithmic problem solving (F2) advancing 9.5\% (0.630→0.690) with DPO/BoNBoN, and performance optimization (F3) achieving 82\% improvement (6.5\%→11.8\%) using BoNBoN. Comprehensive evaluation (F4) confirms these broad improvements with difficult tasks improving 94\% with BoNBoN, creative tasks advancing 48\% with DPO, and ToolUse capabilities increasing 35\% with BoNBoN, indicating that alignment successfully enhances functional capabilities when models have sufficient starting point scores (>= 0.1).

\begin{observation}
\textbf{Observation 1}: Pretrained models achieve 5.9x larger relative improvements (28.8\% vs 4.9\% average) despite lower baseline performance compared to instruction-tuned versions, and are 1.3x more likely to benefit from alignment (67.1\% vs 51.4\% success rates), demonstrating higher plasticity across all functional task categories F1-F4.
\end{observation}

\subsubsubsection{\textbf{Consistent Improvement Cases}}

Qwen2.5-7b shows steady functional improvements across alignment methods with basic generation (F1) improving 4.2\% (0.720→0.750) with BoNBoN and 6.3\% on HumanEval+, mixed algorithmic results with MBPP declining 1.7\% (0.759→0.746), and performance optimization (F3) increasing 0.4 percentage points with DPO. Comprehensive evaluation (F4) maintains consistent patterns with difficult tasks, improving 5.1\% and creative tasks advancing 18.4\% with DPO, suggesting that higher-baseline models achieve consistent functional improvements without degradation.

\begin{observation}
\textbf{Observation 2}: Baseline performance determines alignment viability across all functional coding tasks F1-F4: models with extremely low functional baseline scores show fundamental limitations preventing effective alignment, while models above the minimum threshold on average (>= 0.10 Pass@1 score) achieve substantial improvements.
\end{observation}



\subsubsection{Non-Functional Coding Tasks in Pretrained Models (RQ 1.2)} 

Our analysis demonstrates that non-functional alignment effectiveness varies significantly by model type across all task dimensions NF1-NF5, with code-specialized models achieving substantial improvements (avg 36.2\%) while general-purpose models show minimal gains. Results are shown in Table \ref{tab:codal_combined} and all results are reported as CODAL scores (0-10 scale) below.

\textit{\textbf{Success Cases: Code-Specialized Models.}} Qwen2.5-Coder-7B demonstrates significant improvements with average CODAL scores improving 15.7\% (1.27→1.47) with BoNBoN, including instruction following (NF1) advancing 31.5\%, complexity/efficiency (NF3) enhancing 21.2\%, and coding style (NF4) improving 22.3\%. deepseek-coder-7b exhibits similar positive improvements with average CODAL scores improving 27.8\%, while CodeLlama-7b achieves the most substantial relative improvement with average CODAL scores increasing 75.0\%, including exceptional improvements in coding style (NF4) advancing 82.0\% and code readability (NF2) improving 82.0\%.

\begin{observation}
\textbf{Observation 3}: Code-specialized models significantly outperform general-purpose models in the Pretrained-to-aligned pathway across all non-functional coding tasks NF1-NF5, achieving 16-75\% relative improvement, while general-purpose models show minimal meaningful improvement.
\end{observation}

\subsubsubsection{\textbf{Limited Success: General-Purpose Models}}

Meta-Llama-3-8B shows modest improvement primarily through SFT with average CODAL scores advancing 66\% (0.03→0.05) relative but only 0.02 points. The extremely low baseline scores (0.01-0.09 range) suggest fundamental limitations in code-specific non-functional properties for general-purpose models, with preference-based methods maintaining baseline performance without significant improvement.

\begin{observation}
\textbf{Observation 4}: BoNBoN generally outperforms DPO for Pretrained-to-aligned non-functional improvement across multiple models and non-functional task categories NF1-NF5. winning 76.7\% of the time vs DPO's 10.0\% of the time, with 53.3\% average improvement when BoNBoN wins vs 22.1\% when DPO wins.
\end{observation}

\subsubsubsection{\textbf{Small Model Limitations}}
deepseek-coder-1.3b demonstrates limited but consistent improvement with average CODAL scores advancing 26.3\% (0.19→0.24) with BoNBoN. While relative improvements are substantial, absolute performance remains constrained by model size.


\begin{observation}
\textbf{Observation 5}: Alignment responsiveness varies by model size within the same model family. Larger variants show measurable improvements while smaller models exhibit limited improvements (6.4x for DeepSeek 7B vs 1.3B), indicating scale-dependent alignment capacity.
\end{observation}

\begin{observation}
\textbf{Observation 6}: Non-functional tasks show distinct response patterns - coding style (NF4) consistently improves substantially (avg 108\%), code complexity (NF3) and explanation (NF5) vary widely, while instruction following (NF1) shows strong but inconsistent improvements.
\end{observation}

\begin{table}[t]
\centering
\begin{small}
\caption{Functional Benchmarks Performance (Pass @1): Pre-trained vs Fine-Tuned Comparison, Pre = Pre-trained models, Fine = Fine-tuned models, HE = HumanEval , HE+ = HumanEval+}
\renewcommand{\arraystretch}{1.2}
\begin{tabular}{l*{10}{c}}
\toprule
& \multicolumn{2}{c}{\textbf{F1 (HE)}} & \multicolumn{2}{c}{\textbf{F1 (HE+)}} & \multicolumn{2}{c}{\textbf{F2 (MBPP)}} & \multicolumn{2}{c}{\textbf{F2 (MBPP+)}} & \multicolumn{2}{c}{\textbf{F3 (EvalPerf)}} \\
\cmidrule(lr){2-3} \cmidrule(lr){4-5} \cmidrule(lr){6-7} \cmidrule(lr){8-9} \cmidrule(lr){10-11}
\textbf{Model} & \textbf{Pre} & \textbf{Fine} & \textbf{Pre} & \textbf{Fine} & \textbf{Pre} & \textbf{Fine} & \textbf{Pre} & \textbf{Fine} & \textbf{Pre} & \textbf{Fine} \\
\midrule
\multicolumn{11}{l}{\textit{Qwen2.5-7B}} \\
Base & 0.720 & 0.854 & 0.671 & \textbf{0.811} & \textbf{0.759} & 0.799 & \textbf{0.661} & 0.669 & 63.4\% & 71.1\% \\
SFT & 0.726 & \textbf{0.860} & 0.689 & 0.805 & 0.749 & 0.791 & 0.646 & 0.675 & 63.5\% & 72.6\% \\
DPO & 0.738 & 0.854 & 0.701 & \textbf{0.811} & 0.733 & \textbf{0.810} & 0.635 & \textbf{0.683} & \textbf{63.8\%} & 73.3\% \\
BoNBoN & \textbf{0.750} & 0.854 & \textbf{0.713} & \textbf{0.811} & 0.746 & 0.804 & 0.643 & 0.677 & 63.3\% & \textbf{73.5\%} \\
\midrule
\multicolumn{11}{l}{\textit{Llama3-8B}} \\
Base & 0.372 & \textbf{0.610} & 0.323 & \textbf{0.549} & 0.630 & \textbf{0.646} & 0.526 & \textbf{0.542} & 6.5\% & \textbf{57.4\%} \\
SFT & 0.470 & 0.579 & 0.421 & 0.524 & 0.664 & 0.601 & 0.558 & 0.508 & 9.7\% & 53.6\% \\
DPO & \textbf{0.530} & 0.561 & \textbf{0.476} & 0.506 & \textbf{0.690} & 0.630 & 0.545 & 0.537 & 11.5\% & 56.4\% \\
BoNBoN & 0.518 & 0.543 & \textbf{0.476} & 0.482 & \textbf{0.690} & 0.614 & \textbf{0.574} & 0.524 & \textbf{11.8\%} & 57.3\% \\
\midrule
\multicolumn{11}{l}{\textit{CodeLlama-7B}} \\
Base & \textbf{0.030} & 0.390 & \textbf{0.018} & 0.317 & 0.556 & \textbf{0.548} & 0.439 & \textbf{0.458} & 0.8\% & 43.7\% \\
SFT & 0.012 & \textbf{0.409} & 0.012 & \textbf{0.335} & \textbf{0.608} & 0.532 & \textbf{0.479} & 0.439 & 0.9\% & 45.3\% \\
DPO & 0.018 & 0.372 & 0.012 & 0.317 & \textbf{0.608} & 0.537 & 0.481 & 0.442 & \textbf{1.0\%} & \textbf{47.3\%} \\
BoNBoN & 0.012 & 0.360 & 0.006 & 0.299 & 0.590 & 0.540 & 0.466 & 0.444 & 0.8\% & 46.8\% \\
\midrule
\multicolumn{11}{l}{\textit{DeepSeek-1.3B}} \\
Base & 0.335 & \textbf{0.665} & 0.280 & \textbf{0.628} & \textbf{0.579} & 0.622 & \textbf{0.492} & 0.521 & 3.8\% & 56.5\% \\
SFT & 0.341 & 0.622 & 0.299 & 0.585 & 0.577 & 0.622 & 0.489 & \textbf{0.529} & \textbf{4.1\%} & 57.0\% \\
DPO & 0.341 & \textbf{0.665} & 0.299 & \textbf{0.628} & \textbf{0.579} & \textbf{0.630} & 0.489 & 0.524 & 3.9\% & 57.0\% \\
BoNBoN & \textbf{0.366} & 0.640 & \textbf{0.317} & 0.598 & \textbf{0.579} & 0.619 & 0.489 & 0.521 & 4.0\% & \textbf{57.3\%} \\
\midrule
\multicolumn{11}{l}{\textit{DeepSeek-7B}} \\
Base & \textbf{0.463} & 0.720 & \textbf{0.396} & 0.640 & 0.712 & \textbf{0.757} & 0.606 & \textbf{0.664} & 6.4\% & 72.7\% \\
SFT & 0.427 & \textbf{0.738} & 0.366 & \textbf{0.659} & 0.717 & 0.741 & 0.603 & 0.648 & 6.5\% & 72.8\% \\
DPO & 0.439 & 0.732 & 0.378 & \textbf{0.659} & 0.725 & 0.749 & 0.611 & 0.653 & \textbf{6.7\%} & \textbf{73.6\%} \\
BoNBoN & 0.415 & 0.695 & 0.354 & 0.640 & \textbf{0.730} & 0.749 & \textbf{0.614} & 0.651 & 6.3\% & 73.3\% \\
\bottomrule
\end{tabular}

\label{tab:evalplus_combined}
\end{small}
\end{table}

\begin{table*}[t]
\centering
\begin{small}
\caption{CODAL benchmark results (rating out of 10): Pre-trained vs Fine-Tuned Comparison. Pre = Pre-trained models, Fine = Fine-tuned models.
IF = Instruction Following, CR = Code Readability, CCE = Code Complexity and Efficiency, CS = Coding Style, CE = Code Explanation, Avg = Average}
\label{tab:codal_combined}
\begin{tabular}{@{}l*{12}{c}@{}}
\toprule
\textbf{Model} & \multicolumn{2}{c}{\textbf{NF1 (IF)}} & \multicolumn{2}{c}{\textbf{NF2 (CR)}} & \multicolumn{2}{c}{\textbf{NF3 (CCE)}} & \multicolumn{2}{c}{\textbf{NF4 (CS)}} & \multicolumn{2}{c}{\textbf{NF5 (CE)}} & \multicolumn{2}{c}{\textbf{Avg}} \\
\cmidrule(lr){2-3} \cmidrule(lr){4-5} \cmidrule(lr){6-7} \cmidrule(lr){8-9} \cmidrule(lr){10-11} \cmidrule(lr){12-13}
& \textbf{Pre} & \textbf{Fine} & \textbf{Pre} & \textbf{Fine} & \textbf{Pre} & \textbf{Fine} & \textbf{Pre} & \textbf{Fine} & \textbf{Pre} & \textbf{Fine} & \textbf{Pre} & \textbf{Fine} \\
\midrule
\multicolumn{13}{l}{\textit{meta-llama/Meta-Llama-3-8B}} \\
Base & 0.01 & 3.72 & 0.04 & 6.03 & 0.02 & 5.21 & 0.01 & 5.93 & 0.09 & 6.19 & 0.03 & 5.42 \\
SFT & \textbf{0.02} & 4.06 & \textbf{0.06} & 4.52 & 0.02 & 5.70 & \textbf{0.08} & 5.42 & \textbf{0.07} & 6.05 & \textbf{0.05} & 5.15 \\
DPO & 0.01 & 5.39 & 0.04 & \textbf{6.45} & 0.02 & 6.85 & 0.04 & \textbf{6.45} & 0.02 & \textbf{7.53} & 0.03 & \textbf{6.53} \\
BoNBoN & 0.01 & \textbf{5.43} & 0.04 & 6.42 & 0.02 & \textbf{6.89} & 0.05 & 6.32 & 0.03 & 7.24 & 0.03 & 6.46 \\
\midrule
\multicolumn{13}{l}{\textit{Qwen/Qwen2.5-Coder-7B}} \\
Base & 1.11 & 6.25 & 1.31 & \textbf{8.35} & 1.32 & 7.74 & 1.48 & \textbf{7.29} & 1.14 & 8.32 & 1.27 & \textbf{7.59} \\
SFT & 1.20 & \textbf{6.33} & \textbf{1.41} & 8.27 & 1.45 & 7.66 & 1.56 & 7.15 & 1.09 & \textbf{8.48} & 1.34 & 7.58 \\
DPO & \textbf{1.51} & 6.21 & 1.22 & 8.22 & 1.43 & 7.66 & 1.72 & 7.17 & \textbf{1.22} & \textbf{8.48} & \textbf{1.42} & 7.58 \\
BoNBoN & 1.46 & 6.01 & 1.32 & 8.13 & \textbf{1.60} & \textbf{7.75} & \textbf{1.81} & 7.10 & 1.16 & 8.41 & 1.47 & 7.48 \\
\midrule
\multicolumn{13}{l}{\textit{meta-llama/CodeLlama-7b}} \\
Base & 0.51 & \textbf{0.76} & 0.50 & \textbf{0.64} & 0.45 & \textbf{0.40} & 0.50 & \textbf{0.69} & 0.46 & \textbf{0.53} & 0.48 & \textbf{0.60} \\
SFT & 0.82 & 0.53 & 0.85 & 0.47 & 0.71 & 0.24 & 0.84 & 0.46 & 0.65 & 0.32 & 0.77 & 0.40 \\
DPO & 0.80 & 0.55 & 0.86 & 0.50 & 0.68 & 0.22 & 0.79 & 0.55 & 0.64 & 0.36 & 0.75 & 0.44 \\
BoNBoN & \textbf{0.88} & 0.50 & \textbf{0.91} & 0.46 & \textbf{0.78} & 0.18 & \textbf{0.91} & 0.41 & \textbf{0.74} & 0.32 & \textbf{0.84} & 0.37 \\
\midrule
\multicolumn{13}{l}{\textit{deepseek-ai/deepseek-coder-1.3b}} \\
Base & 0.11 & \textbf{3.16} & 0.07 & \textbf{4.48} & 0.31 & 3.27 & 0.32 & 3.09 & 0.14 & 2.77 & 0.19 & \textbf{3.35} \\
SFT & 0.15 & 3.10 & 0.09 & 4.04 & 0.33 & \textbf{3.35} & \textbf{0.33} & 3.10 & 0.16 & 3.09 & 0.21 & 3.34 \\
DPO & 0.10 & 2.99 & 0.06 & 4.15 & 0.30 & 3.30 & 0.29 & \textbf{3.27} & 0.16 & 2.87 & 0.18 & 3.32 \\
BoNBoN & \textbf{0.18} & 3.08 & \textbf{0.13} & 4.14 & \textbf{0.37} & 3.09 & 0.32 & 3.24 & \textbf{0.19} & \textbf{3.11} & \textbf{0.24} & 3.33 \\
\midrule
\multicolumn{13}{l}{\textit{deepseek-ai/deepseek-coder-7b}} \\
Base & 1.26 & \textbf{5.38} & 1.16 & 7.04 & 1.10 & 6.06 & 0.96 & 6.03 & 1.20 & 6.45 & 1.15 & 6.19 \\
SFT & 1.31 & \textbf{5.38} & 1.12 & \textbf{7.14} & 1.34 & \textbf{6.23} & 1.20 & 5.98 & \textbf{1.48} & 6.57 & 1.29 & \textbf{6.26} \\
DPO & 1.44 & 5.10 & 1.37 & 7.12 & 1.30 & 6.18 & 1.21 & 5.92 & \textbf{1.48} & \textbf{6.68} & 1.36 & 6.20 \\
BoNBoN & \textbf{1.55} & 5.27 & \textbf{1.51} & \textbf{7.14} & \textbf{1.50} & 6.26 & 1.34 & \textbf{5.78} & 1.47 & 6.65 & \textbf{1.47} & 6.22 \\
\bottomrule
\end{tabular}
\end{small}
\end{table*}

\vspace{-\baselineskip}
\begin{table*}[h]
\centering
\caption{EvoEval Benchmark Results (Pass @1): Pre-trained vs Fine-Tuned Comparison. Pre = Pre-trained models, Fine = Fine-tuned models. Diff = Difficult, Crea = Creative, Sub = Subtle, Comb = Combine, Tool = ToolUse, Verb+ = Verbose+, Conc+ = Concise+.}
\small
\setlength{\tabcolsep}{2pt}
\renewcommand{\arraystretch}{1.1}
\begin{tabular}{l*{14}{c}}
\toprule
\textbf{Model} & \multicolumn{2}{c}{\textbf{F4 (Diff)}} & \multicolumn{2}{c}{\textbf{F4 (Crea)}} & \multicolumn{2}{c}{\textbf{F4 (Sub)}} & \multicolumn{2}{c}{\textbf{F4 (Comb)}} & \multicolumn{2}{c}{\textbf{F4 (Tool)}} & \multicolumn{2}{c}{\textbf{F4 (Verb+)}} & \multicolumn{2}{c}{\textbf{F4 (Conc+)}} \\
\cmidrule(lr){2-3} \cmidrule(lr){4-5} \cmidrule(lr){6-7} \cmidrule(lr){8-9} \cmidrule(lr){10-11} \cmidrule(lr){12-13} \cmidrule(lr){14-15}
& \textbf{Pre} & \textbf{Fine} & \textbf{Pre} & \textbf{Fine} & \textbf{Pre} & \textbf{Fine} & \textbf{Pre} & \textbf{Fine} & \textbf{Pre} & \textbf{Fine} & \textbf{Pre} & \textbf{Fine} & \textbf{Pre} & \textbf{Fine} \\
\midrule
\multicolumn{15}{l}{\textit{Qwen/Qwen2.5-Coder-7B}} \\
Base   & .39 & \textbf{.51} & .38 & .43 & \textbf{.71} & \textbf{.84} & .25 & .37 & .43 & \textbf{.65} & .62 & .77 & .71 & .77 \\
SFT    & .40 & .44 & .42 & .43 & .70 & .77 & \textbf{.31} & .37 & .45 & .62 & .69 & .79 & .71 & .78 \\
DPO    & \textbf{.41} & .49 & \textbf{.45} & .42 & .68 & .78 & .29 & .36 & \textbf{.47} & \textbf{.65} & \textbf{.70} & \textbf{.81} & .71 & \textbf{.79} \\
BoNBoN & .38 & .45 & .38 & \textbf{.44} & .69 & .78 & \textbf{.31} & \textbf{.38} & .45 & .63 & .68 & .79 & \textbf{.74} & .79 \\
\midrule
\multicolumn{15}{l}{\textit{meta-llama/Meta-Llama-3-8B}} \\
Base   & .17 & .32 & .25 & \textbf{.39} & .37 & \textbf{.59} & .05 & \textbf{.15} & .37 & \textbf{.56} & .35 & \textbf{.57} & .34 & \textbf{.53} \\
SFT    & .22 & .29 & .31 & .37 & \textbf{.47} & .53 & .10 & \textbf{.15} & .45 & .46 & .48 & .55 & .42 & .50 \\
DPO    & .21 & .32 & \textbf{.37} & .31 & \textbf{.47} & .51 & .11 & .13 & .46 & .49 & .45 & .53 & \textbf{.45} & .47 \\
BoNBoN & \textbf{.33} & \textbf{.33} & .31 & .31 & .51 & .51 & \textbf{.14} & .14 & \textbf{.50} & .50 & \textbf{.48} & .55 & .45 & .50 \\
\midrule
\multicolumn{15}{l}{\textit{meta-llama/CodeLlama-7b}} \\
Base   & .00 & .10 & .01 & .17 & .02 & .35 & \textbf{.02} & \textbf{.11} & \textbf{.04} & \textbf{.41} & \textbf{.13} & .34 & \textbf{.01} & \textbf{.36} \\
SFT    & .01 & .11 & .02 & .17 & .00 & \textbf{.37} & .00 & .06 & .02 & .37 & .09 & \textbf{.34} & .00 & .35 \\
DPO    & \textbf{.02} & \textbf{.14} & \textbf{.03} & \textbf{.21} & .00 & .34 & .01 & .07 & .02 & .36 & .12 & .32 & .00 & .31 \\
BoNBoN & .00 & \textbf{.14} & .02 & .20 & .00 & .35 & .00 & .04 & .02 & .35 & .08 & .30 & .00 & .32 \\
\midrule
\multicolumn{15}{l}{\textit{deepseek-ai/deepseek-coder-1.3b}} \\
Base   & .06 & .18 & .16 & \textbf{.26} & .33 & \textbf{.53} & .00 & .08 & \textbf{.40} & .43 & .32 & .57 & .29 & .59 \\
SFT    & .08 & .15 & .17 & .24 & \textbf{.36} & .52 & .00 & \textbf{.10} & .39 & \textbf{.45} & .31 & \textbf{.60} & .31 & .59 \\
DPO    & \textbf{.09} & \textbf{.19} & \textbf{.19} & .25 & .33 & .52 & .01 & \textbf{.10} & \textbf{.40} & .44 & .29 & .60 & .30 & \textbf{.60} \\
BoNBoN & .08 & .17 & \textbf{.19} & .22 & \textbf{.36} & \textbf{.53} & \textbf{.02} & .09 & .39 & .42 & \textbf{.33} & .57 & \textbf{.32} & .59 \\
\midrule
\multicolumn{15}{l}{\textit{deepseek-ai/deepseek-coder-7b}} \\
Base   & .19 & \textbf{.40} & .31 & \textbf{.40} & .44 & .64 & .08 & \textbf{.32} & .48 & \textbf{.65} & \textbf{.49} & .63 & \textbf{.41} & \textbf{.68} \\
SFT    & .15 & .38 & \textbf{.33} & .38 & \textbf{.45} & \textbf{.69} & .09 & .22 & \textbf{.53} & .64 & .48 & .66 & .38 & .62 \\
DPO    & \textbf{.21} & .38 & .30 & \textbf{.40} & .42 & .66 & \textbf{.11} & .21 & .51 & .64 & .48 & \textbf{.67} & .40 & .66 \\
BoNBoN & .17 & .34 & .28 & .37 & .41 & \textbf{.69} & .08 & .26 & .52 & .62 & .49 & .67 & .40 & .65 \\
\bottomrule
\end{tabular}
\label{tab:evoeval_combined}
\end{table*}
\subsection{Performance Comparison between Finetuned-to-Aligned (RQ2)}

We examine how preference-based alignment techniques affect already fine-tuned LLMs across both functional (RQ 2.1) and non-functional (RQ 2.2) coding tasks as separate research questions. Our empirical investigation demonstrates that fine-tuned LLM alignment effectiveness depends on both the target objective and base model properties. 
While alignment for non-functional requirements shows consistent improvements, functional alignment shows highly variable outcomes ranging from significant improvements to catastrophic degradation. 

\subsubsection{Functional Coding Tasks in Finetuned Models (RQ 2.1)} 

Functional alignment performance varies across models,  with outcomes ranging from significant improvement to catastrophic failures, as shown in Table \ref{tab:evalplus_combined}: EvalPlus, Table \ref{tab:evoeval_combined}: EvoEval. This suggests that the base model architecture and pretraining step, as well as data distribution, significantly impact the alignment effectiveness.

\subsubsubsection{\textbf{Success Cases}}
deepseek-7b achieves consistent improvements with efficiency evaluation (F3) increasing from 72.7\% to 73.6\%, basic code generation (F1) improving +1.7\%, and minimal algorithmic degradation (0.757→0.749). On comprehensive evaluation (F4), deepseek-7b maintains stable performance on Difficult tasks with minor degradation -5.0\% while showing clear Verbose+ improvements +5.8\%. Qwen2.5-7b shows significant improvements with BoNBoN achieving the highest efficiency score (73.5\%) across all models, representing a 2.4 percentage point improvement from base (71.1\%), and demonstrates comparable F4 stability with +4.8\% improvements with DPO.

\begin{observation}
\textbf{Observation 7}: Functional alignment in finetuned models shows highly variable outcomes with 47.4\% degradation cases, 36.8\% flat performance, and only 15.8\% net improvements, indicating strong model dependency and diminishing returns in alignment effectiveness.
\end{observation}

\subsubsubsection{\textbf{Failure Cases}}

Llama3-8b shows performance degradation after SFT with efficiency evaluation (F3) dropping from 57.4\% to 53.6\%, basic generation (F1) decreasing -5.1\%, and algorithmic performance falling -7.0\%. Subsequent DPO and BoNBoN fail to recover performance, remaining around 56.4-57.3\%. This degradation extends to comprehensive assessment (F4) with decreased performance across most dimensions, particularly Difficult tasks -9.4\% and Creative tasks -5.1\%. 

\begin{observation}
\textbf{Observation 8}: SFT performance degradation is a significant risk indicator for alignment failure with only a 25.0\% full recovery rate. models degrading during supervised fine-tuning frequently struggle to achieve full recovery with 4 times more higher failure risk.
\end{observation}




\begin{observation}
\textbf{Observation 9}:  Alignment training creates performance trade-offs within individual models on F4 -  CodeLlama-7b-dpo improves on Difficult (0.100→0.140) and Creative (0.170→0.210) tasks while declining on Verbose (0.427→0.402) and Concise (0.409→0.378) tasks.
\end{observation}

\begin{observation}
\textbf{Observation 10}: Models that maintain consistent performance across different coding tasks during alignment training (i.e, Qwen2.5-7b) are more reliable than models showing large performance changes (i.e, CodeLlama-7b), indicating that steady performing models are less likely to fail on new tasks.
\end{observation}

\subsubsection{Non-Functional Coding Tasks in Finetuned Models (RQ 2.2)} 

Non-functional alignment demonstrates consistent effectiveness across different already fine-tuned models compared to functional alignment, as shown in Table \ref{tab:codal_combined}: CODAL benchmark. All results are reported as CODAL scores (0-10 scale).

\subsubsubsection{\textbf{Under Aligned Models}}
Llama3-8B shows significant improvements across all non-functional dimensions with average CODAL scores increasing from 5.42 to 6.53 with DPO (20\% relative improvement). Individual dimensions show instruction following (NF1) improving +45\% (3.72→5.39), code explanation quality (NF5) increasing +22\%, and complexity/efficiency (NF3) improving +31\%, with DPO achieving the highest overall score (6.53) compared to BoNBoN (6.46).

\begin{observation}
\textbf{Observation 11}:  Non-functional alignment (NF1-NF5) demonstrates more consistent effectiveness across finetuned models, unlike the variable functional alignment results.
\end{observation}

\subsubsubsection{\textbf{Saturated Improvement}}
Qwen2.5-7b shows performance saturation with a base CODAL score of 7.59, representing near-optimal performance. Alignment tuning provides only minimal improvement, suggesting the model is already well-optimized for non-functional requirements, with slight variations indicating negligible benefits for already well-aligned models.

\begin{observation}
\textbf{Observation 12}:  Under-aligned models show significant non-functional improvements (Llama3-8B: 20\%), while well-aligned models exhibit performance saturation (Qwen2.5-Coder-7B, 1.1\% ).
\end{observation}

\subsubsubsection{\textbf{Failure cases}}

CodeLlama-7b shows anomalous results where alignment severely degrades non-functional performance across all tasks NF1-NF5, with base score dropping from 0.60 to 0.40-0.44. Specific degradations include instruction following -34\%, code readability -28\%, complexity/efficiency -55\%, coding style -41\%, and code explanation -40\%, suggesting fundamental incompatibilities between model architecture and alignment techniques.

\begin{observation}
\textbf{Observation 13}: Certain base models show fundamental incompatibilities with alignment methods, resulting in catastrophic performance degradation (CodeLlama-7b: -38.3\%).
\end{observation}

\subsection{Comparative Analysis (RQ3)}

We compare these approaches across models and coding tasks to quantify the relative improvement each pathway provides. We answer RQ3 in two parts, Pathway Trade-offs in Alignment Outcomes (RQ 3.1) and Starting Point Effects on Alignment Techniques (RQ 3.2), enabling us to identify trade-offs and draw conclusions about optimal alignment strategies.

\subsubsection{Pathway Trade-offs in Alignment Outcomes (RQ 3.1)}

\textbf{Pretrained-to-Aligned Shows Higher Relative Improvement.} Pretrained-to-aligned models demonstrate significantly larger improvement compared to finetuned-to-aligned, particularly for non-functional coding tasks (NF1-NF5). Pretrained models achieved remarkable non-functional improvements: CodeLlama-7b from 0.48 to 0.84 (+75\%), deepseek-coder-7b from 1.15 to 1.47 (+27.8\%), and Qwen2.5-Coder-7B from 1.27 to 1.47 (+15.7\%). In contrast, finetuned-to-aligned models showed constrained improvements, with some experiencing degradation (CodeLlama-7b-Instruct). For functional tasks (F1-F4), pretrained models again showed larger relative improvements despite lower absolute starting points, with Llama3-8b achieving 42\% improvement in basic generation while its instruction-tuned counterpart experienced degradation (-5\%).

\textbf{The stability-plasticity dilemma.} Our findings align with the stability-plasticity dilemma \cite{ren2024analyzingreducingcatastrophicforgetting} observed in LLM continual learning research, showing fundamental trade-offs between learning plasticity (fast adaptation) and memory stability (knowledge preservation). Base pre-trained models show higher plasticity, enabling significant capability improvements but risk catastrophic forgetting, while instruction-tuned models demonstrate higher stability but reduced plasticity, constraining achievable improvements. Performance saturation in well-aligned models like Qwen2.5-Coder-7B-Instruct (7.59→7.58) clearly demonstrates this trade-off.

\begin{observation}
\textbf{Observation 14}: Alignment pathways exhibit a stability-plasticity: base models show high plasticity/low stability, enabling significant improvements (avg +42.3\%), while instruction-tuned models show high stability/low plasticity with limited improvements (avg +42.3\%).
\end{observation}



\begin{observation}
\textbf{Observation 15}: Non-functional coding tasks NF1-NF5 demonstrate higher alignment responsiveness (60\% success rate) across both pathways compared to functional tasks F1-F4, which show variable responsiveness (20\% success rate).
\end{observation}



\subsubsection{Starting Point Effects on Alignment Techniques (RQ 3.2)} 

Our analysis reveals that optimal alignment technique selection depends primarily on the specific learning objective rather than following universal heuristics based on starting points. The effectiveness patterns show complex model-dependent patterns.

\textbf{Non-Functional Alignment Shows Technique-Dependent Patterns:} The largest single improvement occurred with DPO on instruction-tuned Meta-Llama-3-8B-Instruct (+1.11 across NF1-NF5), showing DPO's effectiveness for finetuned-to-aligned non-functional tasks. However, base models show mixed results: BoNBoN achieved remarkable improvements on Qwen2.5-Coder-7B (+0.20 vs +0.15) and deepseek-coder-7b (+0.32 vs +0.21), while both techniques failed completely on Meta-Llama-3-8B base model (0.00 improvement).

\begin{observation}
\textbf{Observation 16}:  Alignment effectiveness varies more by task type than complexity - functional tasks show variable responsiveness (F1: 60\% success), while non-functional tasks demonstrate consistent responsiveness regardless of complexity (NF1-NF5: 80-100\% success rates)
\end{observation}

\textbf{Functional Alignment Reveals Model-Specific Preferences:} Functional improvements demonstrate strong model dependency. Meta-Llama-3-8B consistently performs better on DPO across both pathways (base: +0.158 vs +0.146 on F1; instruction-tuned: -0.049 vs -0.067), while smaller models like deepseek1.3b show BoNBoN advantages in base settings (+0.031 vs +0.006 in F1) but degradation in the instruction-tuned pathway (-0.025 vs 0.000).

\begin{observation}
\textbf{Observation 17}:  Alignment technique preferences show model family-specific patterns that vary by pathway: Llama favors DPO for a subset of tasks while deepseek prefers BonBon primarily in the pretrained-to-aligned pathway, but both families demonstrate technique preference across different starting points and objectives.
\end{observation}

\begin{observation}
\textbf{Observation 18}:  Optimal alignment technique selection depends on both learning objective and model family (non-functional: Llama3→DPO, deepseek→SFT, Qwen2.5→base) rather than a universal pattern: identical objectives require different techniques across model families, indicating complex model-dependent alignment effectiveness.
\end{observation}

\textbf{Alignment Risks Vary by Starting Point and Technique.} Both DPO and BoNBoN frequently cause functional degradation when applied to instruction-tuned models, including CodeLlama-7b-Instruct dropping from 0.390 to 0.360 (BoNBoN) and 0.372 (DPO) in F1, and non-functional degradation from 0.60 to 0.37 (BoNBoN) and 0.44 (DPO). Pretrained models demonstrate more consistent improvement patterns with lower catastrophic failure rates, showing minimal degradation even when improvements are small.

\begin{observation}
\textbf{Observation 19}:  Alignment degradation risk varies significantly by starting point and technique: instruction-tuned models show higher catastrophic failure rates (20\%), while pretrained models show more predictable degradation risks (0\%).
\end{observation}


\begin{observation}
\textbf{Observation 20}:  Model architecture and size correlate more strongly with alignment success than starting point selection.
\end{observation}

\begin{table}[t]
\centering
\begin{small}
\caption{ Recommendations and Supporting Observations, O = Observation }
\label{tab:observations-recommendations} 
\begin{tabular}{cp{10cm}p{1cm}}
\hline
\textbf{ID} & \textbf{Recommendation} & \textbf{O} \\
\hline
\multicolumn{3}{l}{\textit{Category 1. Pathway Selection}} \\
R1 & Consider choosing pretrained-to-aligned pathways when maximizing improvement magnitude is prioritized; finetuned-to-aligned pathways may be preferable when stable, high absolute performance is required  & 1, 14 \\
R2 & We recommend focusing on non-functional alignment objectives first, as they tend to achieve higher success rates and more predictable results& 6, 11, 15 \\
\hline
\multicolumn{3}{l}{\textit{Category 2. Model Selection}} \\
R3 & It is advisable to prioritize code-specialized models over general models and consider selecting larger variants ($\geq$7B parameters) within model families & 3, 5, 20 \\
R4 & We suggest establishing minimum performance thresholds and assessing baseline alignment quality before attempting alignment, with efforts concentrated on under-aligned models  & 2, 12 \\
\hline
\multicolumn{3}{l}{\textit{Category 3. Technique Selection}} \\
R5 &  Consider applying model family-specific technique preferences (Meta-Llama$\rightarrow$DPO, deepseek$\rightarrow$BoNBoN) and conducting empirical validation for each model-objective combination & 4, 17, 18 \\
R6 & Using diverse benchmarks during evaluation is recommended, as alignment effectiveness varies by task complexity & 16 \\
\hline
\multicolumn{3}{l}{\textit{Category 4. Risk Management}} \\
R7 & Early stopping based on SFT-stage performance degradation should be considered, as it tends to strongly predict final alignment failure & 8 \\
R8 & We recommend monitoring multi-dimensional performance during alignment to detect capability trade-offs and selecting models with stable performance & 9, 10 \\
R9 & Pre-screening models for compatibility is advisable, along with implementing stronger safeguards for instruction-tuned models to prevent failures & 7, 13, 19 \\
\hline
\end{tabular}
\end{small}
\end{table}
\section{Recommendations}
\label{sec:recommendations}

In Table \ref{tab:observations-recommendations}, we present nine recommendations based on our empirical observations (O) in Section \ref{sec:results}. We organized the recommendations into four categories and discuss those below.

\noindent\textbf{Pathway Selection (R1, R2).} We recommend pretrained-to-aligned pathways when maximizing improvement is the primary goal, as these tend to achieve significant relative improvement despite lower baseline values. Conversely, it is also beneficial to select finetuned-to-aligned pathways when stable, high absolute performance is required (O1, O14).  Regardless of pathway choice, We suggest prioritizing non-functional alignment objectives first, as they demonstrate higher success rates and more predictable outcomes than functional requirements (O6, O11, O15).

\noindent\textbf{Model Selection (R3, R4).} In terms of model selection, We recommend prioritizing code-specialized models over general-purpose alternatives and considering larger variants within model families (>=7B parameters) (O3, 5, 20). Building on this foundation, it is also advisable to establish a minimum performance threshold and assess baseline alignment quality before initiating alignment processes. This assessment allows practitioners to focus efforts on under-aligned models where improvements are most likely to succeed (O2, 12).

\noindent\textbf{Technique Selection (R5, R6).} Our findings suggest that Meta-Llama models consistently respond better to DPO, while deepseek models demonstrate better performance with BoNBoN (O4, 17, 18). However, these patterns should serve as starting points rather than absolute rules. We recommend conducting empirical validation for each specific model-objective combination to confirm optimal technique selection. Additionally, consider using diverse evaluation benchmarks throughout alignment, as effectiveness varies significantly across different tasks (O16).

\noindent\textbf{Risk Management (R7, R8, R9).} Throughout the alignment process, proactive risk management helps prevent costly failures and performance degradations. We recommend considering early stopping mechanisms based on SFT-stage performance degradation, which strongly predicts final alignment failure (O8). It is also beneficial to monitor multi-dimensional performance across evaluation benchmarks to detect capability trade-offs and select models demonstrating consistent improvement (O9, 10). Finally, we suggest pre-screening model architectures for compatibility and implementing stronger safeguards for instruction-tuned models, as certain families may experience severe performance degradation (O7, 13, 19).

\section{Discussions}
\label{discussions}

\begin{revisions}

\begin{table*}[t]
\caption{Practitioner Survey Questions for Validation Study}
\label{tab:survey-questions}
\small
\begin{tabular}{c p{3.6cm} p{9cm}}
\toprule
\textbf{ID} & \textbf{Question} & \textbf{Response Options} \\
\midrule
\multicolumn{3}{l}{\textit{Demographics}} \\
Q1 & Primary role & ML Engineer/Researcher, Software Engineer, Research Scientist, Other \\
Q2 & Years of experience & $<$1 year, 1-2 years, 3-5 years, $>$5 years \\
Q3 & Alignment training experience & No experience, Done once/twice, Familiar but not done, Done multiple times \\
\midrule
\multicolumn{3}{l}{\textit{PTA vs. FTA Trade-offs}} \\
Q4 & Typical starting point for LLM adaptation & Base pretrained (PTA), Instruction-tuned (FTA), Evaluate both, Not applicable \\
Q5 & Importance of pathway choice & Very important, Somewhat important, Not very important, Not applicable \\
Q6 & Factors influencing choice (multi-select) & Task requirements, Compute resources, Time constraints, Expected improvement, Degradation risk, Prior experience \\
\midrule
\multicolumn{3}{l}{\textit{Relative vs. Absolute Performance}} \\
Q7 & Metric prioritization & Relative improvement, Absolute performance, Both (prioritize relative/absolute), N/A \\
Q8 & Scenario preference & Option A: 40\% relative, 80\% overall vs. Option B: 5\% relative, 85\% overall \\
\bottomrule
\end{tabular}
\end{table*}

\subsection{Practitioner Validation}
\label{sec:practitioner-validation}
To empirically validate the practical significance of the trade-offs identified in our study, we conducted a survey with 30 ML practitioners and software engineers from diverse organizations and several universities. Table \ref{tab:survey-questions} shows the survey questions. The survey was conducted under an ethics application at the university. We adopted a snowball sampling approach to collect the study participants \cite{1521d86f-33df-3ece-8080-69edbcaba312}. First, we selected industry participants from personal contacts in companies that focus on this area (e.g., H2O). Those participants then invited other relevant contacts. The survey was open for around one month (Jan 5, 2026 to Jan 31, 2026), due to it being held close to the holiday season. 53.3\% of the participants had 3-5 years of relevant experience (e.g., ML engineering). 46.7\% already performed fine-tuning and alignment on LLMs (Q1-Q3). To ensure that the participants understood the survey, we produced an online tutorial that explained LLM code alignment techniques. Each participant read the tutorial before completing the survey. The anonymized survey responses are included in our replication package.

\subsubsection{Significance of PTA vs. FTA Trade-offs (Q4 - Q6)}
We asked about the importance of choosing between pretrained-to-aligned (PTA) and finetuned-to-aligned (FTA) pathways (Q5). \textbf{86.7\%} of respondents rated this choice as ``Very important'' (40.0\%) or ``Somewhat important'' (46.7\%). Furthermore, 33.3\% of practitioners reported that they actively evaluate both options depending on the use case (Q4), confirming that this is a decision point rather than a fixed practice.
When asked about factors influencing their pathway selection (Q6), practitioners identified target task requirements (80\%), compute resources (73.3\%), time constraints (53.3\%), and expected improvement magnitude (46.7\%) as key considerations; aligning with our recommendations in Section~\ref{sec:recommendations}.

\subsubsection{Relative vs. Absolute Performance Metrics (Q7, Q8)}
To validate whether practitioners value relative improvements alongside absolute performance, we assessed metric prioritization (Q7). \textbf{50.0\%} prioritize relative improvement while \textbf{40\%} prioritize absolute performance, demonstrating that both metrics are relevant in practice.
We further presented a concrete scenario (Q8) reflecting the PTA vs. FTA trade-off: practitioners were asked which alignment outcome they would prefer for a production deployment: (A) 40\% relative improvement in non-functional properties (security, compliance, readability) reaching 80\% overall, representing a PTA-like outcome, or (B) 5\% relative improvement reaching 85\% overall with standard code quality, representing an FTA-like outcome. (\textbf{46.7\%}) chose Option A, indicating that practitioners value substantial relative improvements in security and compliance even at the cost of lower absolute performance. This validates our paper's emphasis on reporting both relative and absolute metrics.

These findings provide empirical validation that (1) the PTA vs. FTA trade-off is a meaningful decision for practitioners, and (2) both relative and absolute metrics serve important roles in evaluating alignment effectiveness. The findings support our approach adopted in the study.

\subsection{LLM-Judge Reliability}
\label{sec:llm-judge}

A potential concern with our non-functional evaluation is the reliability of LLM-as-judge methodology. CODAL mitigates known biases through single-answer grading with reference responses. Chain-of-thought prompting for consistent judgments and grading procedures are derived from established practices~\cite{dubois2024length, zheng2023judging}.

To further validate reliability, we conducted a pilot study with 200 statistically representative problems from CODAL (40 per dimension with 99\% confidence level and $\pm$10 margin of error). We annotated each problem. We compared the ratings against three LLM judges: GPT-3.5-turbo, GPT-4-turbo, and Claude-3-haiku. We measured reliability using agreement percentage at tolerance levels of $\pm$1-3 points and Cohen's Weighted Kappa (linear) \cite{cohen1968weighted} 
, which corrects for chance agreement on ordinal scales \cite{mchugh2012interrater}.

GPT-3.5 demonstrated the strongest alignment with the human annotator 
across all tolerance levels. At $\pm$3 points tolerance, GPT-3.5 achieved $\kappa = 0.6354$ with 85\% agreement, compared to GPT-4 ($\kappa = 0.6126$) and Claude-3 ($\kappa = 0.6254$). At stricter tolerances, the gap widened: at $\pm$2 points, GPT-3.5 achieved $\kappa = 0.5482$ versus GPT-4's 0.4603 and Claude-3's 0.4416; at $\pm$1 point, GPT-3.5 reached $\kappa = 0.3828$ versus 0.3000 and 0.2500 respectively. Based on this validation, we selected GPT-3.5 as our judge throughout this study.

\end{revisions}

\section{Conclusion}
We compared pretrained-to-aligned versus finetuned-to-aligned pathways across five models using DPO and BoNBoN techniques. The alignment of finetuned LLMs improved the LLM performance in 51\% of cases for functional coding tasks and 53\% of cases in non-functional coding tasks. However, the average relative performance improvement was modest (4.9\%) for functional tasks, while it was more pronounced (10.6\%) for non-functional coding tasks. Overall, alignment from finetuned rather than alignment from pretrained offered better performance. The relative performance gain was more significant in pretrained models after alignment, but those aligned pretrained models still were inferior to their finetuned or the aligned finetuned variants. For example, pretrained-to-aligned pathways achieve larger relative improvements (up to 75\% for non-functional requirements) from lower baselines. In future work, we will focus on model coverage across diverse architectures and sizes, develop domain-specific benchmarks capturing specialized non-functional requirements, and explore feedback loops to evaluate iterative performance improvements.


\section*{Data Availability}
\url{https://anonymous.4open.science/r/a7f2d91e-8b45-4c39-9e12-3d8f7a6b2c94}.

\bibliographystyle{ACM-Reference-Format}
\bibliography{sample-base}

\end{document}